\documentstyle[twoside]{article}

\catcode`\@=11
\long\def\@makefntext#1{
\protect\noindent \hbox to 3.2pt {\hskip-.9pt  
$^{{\eightrm\@thefnmark}}$\hfil}#1\hfill}               

\def\@makefnmark{\hbox to 0pt{$^{\@thefnmark}$\hss}}    
        
\def\ps@myheadings{\let\@mkboth\@gobbletwo
\def\@oddhead{\hbox{}
\rightmark\hfil\eightrm\thepage}   
\def\@oddfoot{}\def\@evenhead{\eightrm\thepage\hfil
\leftmark\hbox{}}\def\@evenfoot{}
\def\sectionmark##1{}\def\subsectionmark##1{}}

\oddsidemargin=\evensidemargin
\newcounter{sectionc}\newcounter{subsectionc}\newcounter{subsubsectionc}
\renewcommand{\section}[1] {\vspace{12pt}\addtocounter{sectionc}{1} 
\setcounter{subsectionc}{0}\setcounter{subsubsectionc}{0}\noindent 
        {\tenbf\thesectionc. #1}\par\vspace{5pt}}
\renewcommand{\subsection}[1] {\vspace{12pt}\addtocounter{subsectionc}{1} 
        \setcounter{subsubsectionc}{0}\noindent 
        {\bf\thesectionc.\thesubsectionc. {\kern1pt \bfit #1}}\par\vspace{5pt}}
\renewcommand{\subsubsection}[1] {\vspace{12pt}\addtocounter{subsubsectionc}{1}
        \noindent{\tenrm\thesectionc.\thesubsectionc.\thesubsubsectionc.
        {\kern1pt \tenit #1}}\par\vspace{5pt}}
\newcommand{\nonumsection}[1] {\vspace{12pt}\noindent{\tenbf #1}
        \par\vspace{5pt}}

\newcounter{appendixc}
\newcounter{subappendixc}[appendixc]
\newcounter{subsubappendixc}[subappendixc]
\renewcommand{\thesubappendixc}{\Alph{appendixc}.\arabic{subappendixc}}
\renewcommand{\thesubsubappendixc}
        {\Alph{appendixc}.\arabic{subappendixc}.\arabic{subsubappendixc}}

\renewcommand{\appendix}[1] {\vspace{12pt}
        \refstepcounter{appendixc}
        \setcounter{figure}{0}
        \setcounter{table}{0}
        \setcounter{lemma}{0}
        \setcounter{theorem}{0}
        \setcounter{corollary}{0}
        \setcounter{definition}{0}
        \setcounter{equation}{0}
        \renewcommand{\thefigure}{\Alph{appendixc}.\arabic{figure}}
        \renewcommand{\thetable}{\Alph{appendixc}.\arabic{table}}
        \renewcommand{\theappendixc}{\Alph{appendixc}}
        \renewcommand{\thelemma}{\Alph{appendixc}.\arabic{lemma}}
        \renewcommand{\thetheorem}{\Alph{appendixc}.\arabic{theorem}}
        \renewcommand{\thedefinition}{\Alph{appendixc}.\arabic{definition}}
        \renewcommand{\thecorollary}{\Alph{appendixc}.\arabic{corollary}}
        \renewcommand{\theequation}{\Alph{appendixc}.\arabic{equation}}
        \noindent{\tenbf Appendix \theappendixc #1}\par\vspace{5pt}}
\newcommand{\subappendix}[1] {\vspace{12pt}
        \refstepcounter{subappendixc}
        \noindent{\bf Appendix \thesubappendixc. {\kern1pt \bfit #1}}
        \par\vspace{5pt}}
\newcommand{\subsubappendix}[1] {\vspace{12pt}
        \refstepcounter{subsubappendixc}
        \noindent{\rm Appendix \thesubsubappendixc. {\kern1pt \tenit #1}}
        \par\vspace{5pt}}

\newcommand{\textlineskip}{\baselineskip=13pt}
\newcommand{\smalllineskip}{\baselineskip=10pt}

\def\eightcirc{
\begin{picture}(0,0)
\put(4.4,1.8){\circle{6.5}}
\end{picture}}
\def\eightcopyright{\eightcirc\kern2.7pt\hbox{\eightrm c}}

\def\abstracts#1#2#3{{
        \centering{\begin{minipage}{4.5in}\baselineskip=10pt\footnotesize
        \parindent=0pt #1\par 
        \parindent=15pt #2\par
        \parindent=15pt #3
        \end{minipage}}\par}} 

\renewenvironment{thebibliography}[1]
        {\frenchspacing
         \ninerm\baselineskip=11pt
         \begin{list}{\arabic{enumi}.}
        {\usecounter{enumi}\setlength{\parsep}{0pt}
         \setlength{\leftmargin 12.7pt}{\rightmargin 0pt} 
         \setlength{\itemsep}{0pt} \settowidth
        {\labelwidth}{#1.}\sloppy}}{\end{list}}

\newcounter{itemlistc}
\newcounter{romanlistc}
\newcounter{alphlistc}
\newcounter{arabiclistc}

\newcommand{\fcaption}[1]{
        \refstepcounter{figure}
        \setbox\@tempboxa = \hbox{\footnotesize Fig.~\thefigure. #1}
        \ifdim \wd\@tempboxa > 5in
           {\begin{center}
        \parbox{5in}{\footnotesize\smalllineskip Fig.~\thefigure. #1}
            \end{center}}
        \else
             {\begin{center}
             {\footnotesize Fig.~\thefigure. #1}
              \end{center}}
        \fi}

\newcommand{\tcaption}[1]{
        \refstepcounter{table}
        \setbox\@tempboxa = \hbox{\footnotesize Table~\thetable. #1}
        \ifdim \wd\@tempboxa > 5in
           {\begin{center}
        \parbox{5in}{\footnotesize\smalllineskip Table~\thetable. #1}
            \end{center}}
        \else
             {\begin{center}
             {\footnotesize Table~\thetable. #1}
              \end{center}}
        \fi}

\def\@citex[#1]#2{\if@filesw\immediate\write\@auxout
        {\string\citation{#2}}\fi
\def\@citea{}\@cite{\@for\@citeb:=#2\do
        {\@citea\def\@citea{,}\@ifundefined
        {b@\@citeb}{{\bf ?}\@warning
        {Citation `\@citeb' on page \thepage \space undefined}}
        {\csname b@\@citeb\endcsname}}}{#1}}

\newif\if@cghi
\def\cite{\@cghitrue\@ifnextchar [{\@tempswatrue
        \@citex}{\@tempswafalse\@citex[]}}
\def\citelow{\@cghifalse\@ifnextchar [{\@tempswatrue
        \@citex}{\@tempswafalse\@citex[]}}
\def\@cite#1#2{{$\null^{#1}$\if@tempswa\typeout
        {IJCGA warning: optional citation argument 
        ignored: `#2'} \fi}}

\def\pmb#1{\setbox0=\hbox{#1}
        \kern-.025em\copy0\kern-\wd0
        \kern.05em\copy0\kern-\wd0
        \kern-.025em\raise.0433em\box0}

\def\fnt#1#2{\footnotetext{\kern-.3em
        {$^{\mbox{\scriptsize #1}}$}{#2}}}

\def\fpage#1{\begingroup
\voffset=.3in
\thispagestyle{empty}\begin{table}[b]\centerline{\footnotesize #1}
        \end{table}\endgroup}

\font\tenrm=cmr10
\font\tenit=cmti10 
\font\tenbf=cmbx10
\font\bfit=cmbxti10 at 10pt
\font\ninerm=cmr9

\font\eightrm=cmr8

\def\qed{\hbox{${\vcenter{\vbox{                        
   \hrule height 0.4pt\hbox{\vrule width 0.4pt height 6pt
   \kern5pt\vrule width 0.4pt}\hrule height 0.4pt}}}$}}

\def\bffrac#1/#2{\leavevmode\kern.1em
\raise.5ex\hbox{$\bf\scriptstyle #1$}\kern-.1em
{\bf/}\kern-.15em\lower.25ex\hbox{$\bf\scriptstyle #2$}}

\catcode`\@=11
\def\eqnarray{\stepcounter{equation}\let\@currentlabel=\theequation
\global\@eqnswtrue
\global\@eqcnt\z@\tabskip\@centering\let\\=\@eqncr
$$\halign to \displaywidth\bgroup\@eqnsel\hskip\@centering
  $\displaystyle\tabskip\z@{##}$&\global\@eqcnt\@ne
  \hfil${##}$\hfil
  &\global\@eqcnt\tw@ 
$\displaystyle\tabskip\z@{##}$\hfil
   \tabskip\@centering&\llap{##}\tabskip\z@\cr}
\catcode`\@=12
\catcode`\@=11
\def\lowcite{\def\@cite##1##2{{${##1}$\if@tempswa\typeout
{IJCGA warning: optional citation argument
        ignored: `##2'} \fi}}}
\catcode`\@=12
\begin{document}

\normalsize\textlineskip
\thispagestyle{empty}
\setcounter{page}{1}

%
%
\fpage{1}
\hbox{}
\vskip -2.32cm \rightline{
\vbox{\rm
\halign{&# \hfil\cr
&ANL-HEP-CP-96-79\cr
&September 1996\cr}
}}\vskip 2.00cm%
\centerline{\bf LATTICE CALCULATION OF QUARKONIUM}
\vspace*{0.035truein}
\centerline{\bf DECAY MATRIX ELEMENTS\footnote{Talk presented by
G.T.~Bodwin at the Quarkonium Physics Workshop, University of Illinois, 
Chicago, June 13--15, 1996.  This work has been reported previously in 
Ref.~\lowcite\cite{bks}.}}
\vspace*{0.37truein}
\centerline{\footnotesize G.T.~BODWIN AND D.K.~SINCLAIR}
\vspace*{0.015truein}
\centerline{\footnotesize\it HEP Division, Argonne National Laboratory,
9700 South Cass Avenue}
\baselineskip=10pt
\centerline{\footnotesize\it Argonne, Illinois 60517, USA}
\vspace*{10pt}
\centerline{\footnotesize S.~KIM}
\vspace*{0.015truein}
\centerline{\footnotesize\it Center for Theoretical Physics, Seoul
National University}
\baselineskip=10pt
\centerline{\footnotesize\it Seoul, Korea}
\vspace*{0.225truein}

\vspace*{0.21truein}
\abstracts{We calculate the NRQCD matrix elements for the decays
of the lowest-lying S- and P-wave states of charmonium and bottomonium
in quenched lattice QCD. 
We also compute the one-loop relations between the lattice and continuum
matrix elements.}{}{} 

\vspace*{1pt}\textlineskip      
\section{Heavy-Quarkonium Formalism}
\vspace*{-0.5pt}
\noindent

Heavy-quarkonium systems are nonrelativistic.  In the CM frame, the
average quark velocity $v$ satisfies $v^2 \ll 1$, where $v^2 \approx
0.3$ for charmonium, and $v^2 \approx 0.1$ for bottomonium.  This fact
allows one to describe heavy-quarkonium systems conveniently in terms of
the effective field theory Nonrelativistic QCD
(NRQCD).\cite{caswell-lepage} NRQCD accurately describes processes in
which $p_Q< M_Q$.  Its utility stems from the fact that it can be used
to decouple short-distance ($\sim 1/M_Q$) processes from long-distance
($\sim\hbox{quarkonium size}\sim 1/(M_Qv)$) processes. 

$Q\overline Q$ annihilation occurs at a distance of order $1/M_Q$, so
NRQCD does {\it not} describe the details of that process.  In NRQCD,
the short-distance part of the amplitude for $Q\overline Q\rightarrow
\hbox{light hadrons}\rightarrow Q\overline Q$ is pointlike, and the
entire amplitude is described by a four-fermion interaction. The
quarkonium total annihilation rate is proportional to the imaginary part
of the matrix element in the quarkonium state of the appropriate
four-fermion operator. 

Coefficients of the four-fermion operators are determined by matching
matrix elements in NRQCD to those in full QCD. The coefficients are
short-distance quantities and, hence, are calculable in QCD perturbation
theory.  They are proportional to the IR-finite parts of the parton-level
annihilation rates. 

\subsection{Factorization theorems}
\noindent

Using these ideas, Bodwin, Braaten, and Lepage\cite{bbl} have shown that
a quarkonium decay rate can be written as a sum of terms. Each term is
the product of a long-distance matrix element of a four-fermion operator
in the quarkonium state with a short-distance coefficient. For example,
decay rates for S-wave quarkonia through next-to-leading order in $v^2$
are given by 
\begin{eqnarray}
\Gamma(^{2s+1}S_J\rightarrow X) & = & {\cal G}_1(^{2s+1}S_J)\,2\,
					{\rm Im\,}f_1
                                      (^{2s+1}S_J)/M_Q^2 
                                      \nonumber \\
                                & + & {\cal F}_1(^{2s+1}S_J)\,2\,{\rm Im\,}g_1
                                      (^{2s+1}S_J)/M_Q^4.
\label{s-wave}
\end{eqnarray}
Decay rates for P-wave quarkonia, to the lowest non-trivial order in 
$v^2$, are given by 
\begin{eqnarray}
\Gamma(^{2s+1}P_J\rightarrow X) & = & {\cal H}_1(^{2s+1}P_J)\,2\,{\rm Im\,}f_1
                                      (^{2s+1}P_J)/M_Q^4
                                      \nonumber \\
                                & + & {\cal H}_8(^{2s+1}P_J)\,2\,{\rm Im\,}f_8
                                      (^{2s+1}S_J)/M_Q^2.
\label{p-wave}
\end{eqnarray}
The $f$'s and $g$'s are the short-distance coefficients and are
proportional to the rates for the annihilation of a $Q\bar Q$ pair from
the $^{2s+1}L_J$ state. ${\cal G}_1$, ${\cal F}_1$, ${\cal H}_1$, and
${\cal H}_8$ are the long-distance matrix elements, which we calculate
in this paper. The subscripts $1$ and $8$ indicate the color state of
the $Q\bar Q$ pair. (An octet state, in lowest order in $v$, consists of
a $Q\bar{Q}g$ state.) 

\subsection{Matrix elements}
\noindent

The long-distance matrix elements are defined by
\begin{eqnarray}
{\cal G}_1 & = & 
\langle^1S_0|\psi^{\dagger}\chi\chi^{\dagger}\psi|^1S_0\rangle,\nonumber\\
{\cal F}_1 & = & \langle^1S_0|\psi^{\dagger}\chi\chi^{\dagger} 
(\frac{-i}{2}\stackrel{\leftrightarrow}{\bf D})^2\psi 
|^1S_0\rangle,\nonumber \\
{\cal H}_1 & = &
\langle ^1P_1|\psi^{\dagger}(i/2)\stackrel{\leftrightarrow}{\bf D}
\chi.\chi^{\dagger}(i/2)\stackrel{\leftrightarrow}{\bf 
D}\psi|^1P_1\rangle,\nonumber \\
{\cal H}_8 & = & \langle 
^1P_1|\psi^{\dagger}T^a\chi\chi^{\dagger}T^a\psi|^1P_1\rangle.
\label{matrix-elements}
\end{eqnarray}                                                   
Here, $\psi$ and $\chi^\dagger$ are two-component spinors that
annihilate a heavy quark and a heavy antiquark, respectively. 

The ${\cal G}_1$ and ${\cal H}_1$ terms in the decay rates
(\ref{s-wave}) and (\ref{p-wave}) appear in the conventional,
color-singlet model.\cite{color-singlet-model} 
For the color-singlet matrix elements we can, to good approximation,
take only the vacuum in the intermediate state.\cite{bbl} Then we have 
\begin{eqnarray}
{\cal G}_1&&= |\langle^1S_0|\psi^{\dagger}\chi|0\rangle|^2 \;(1+{\cal 
O}(v^4)),\nonumber\\
{\cal F}_1&&=\langle^1S_0|\psi^{\dagger}\chi|0\rangle
\langle 0|\chi^{\dagger} 
(\frac{-i}{2}\stackrel{\leftrightarrow}{\bf D})^2\psi |^1S_0\rangle 
\;(1+{\cal O}(v^4)),\nonumber\\
{\cal H}_1&&=|\langle^1P_1|\psi^{\dagger}\frac{-i}{2}\stackrel{\leftrightarrow}
                        {\bf D}\chi|0\rangle|^2 \;(1+{\cal 
O}(v^4)).
\end{eqnarray}
It then follows that, in the vacuum-saturation approximation,
\begin{eqnarray}{\cal G}_1 &&\approx
\frac{3}{2\pi}|R_S(0)|^2,\nonumber\\
{\cal H}_1 &&\approx \frac{9}{2\pi}|R'_P(0)|^2,
\label{vacuum-sat}
\end{eqnarray}
where $R(0)$ is the radial wavefunction at the origin, and $R'(0)$ is
the derivative of the radial wavefunction at the origin. The
matrix-elements (\ref{matrix-elements}) define a regularized $R(0)$ and
a regularized $R'(0)$ in QCD.  The ratio ${\cal F}_1/{\cal G}_1$
measures the average of $\vec p^{\,2}$ in the quarkonium state. 

The term in the P-wave decay rate (\ref{p-wave}) that is proportional
to ${\cal H}_8$ is absent in the color-singlet model.  The matrix
element ${\cal H}_8$ is proportional to the probability to find a
$Q\bar{Q}g$ component in P-wave quarkonium, with the $Q\bar Q$ pair in a
relative S-wave, color-octet state.  It is perhaps the most interesting
of the matrix elements that we measure, since it corresponds to a
field-theoretic effect of QCD and is, therefore, inaccessible through
any potential model of quarkonium. 

\vspace*{1pt}\textlineskip
\section{Lattice Measurement of the Matrix Elements}
\vspace*{-0.5pt}
\noindent

\subsection{Euclidean-space formalism}
\noindent

In Euclidean space, quarkonium two-point functions decay exponentially:
\begin{equation}
\lim_{T\rightarrow\infty}
\langle M(T)\tilde M^\dagger(0)\rangle =\lim_{T\rightarrow\infty}
\langle 0|Me^{-HT}\tilde M^\dagger|0\rangle
=\langle 0|M|l\rangle 
e^{-E_lT}\langle l|\tilde M^\dagger|0\rangle.
\label{2point}
\end{equation}
Here, $M$ and $\tilde M$ are quarkonium sources, and $|l\rangle $ is the
lowest-lying state with the quarkonium quantum numbers. We deduce from
(\ref{2point}) that, in the vacuum-saturation approximation, the
quarkonium matrix elements are proportional to the coefficients of
$e^{-E_lT}$ in the appropriate two-point functions. 

Similarly, a three-point function involving the four-fermion 
operator ${\cal O}$ has the behavior
\begin{eqnarray}
\lim_{T,T'\rightarrow\infty}
\langle M(T+T')\;{\cal O}(T)\;M^\dagger(0)\rangle
&=&\langle 0|M|l\rangle e^{-E_lT'}\langle l|{\cal 
O}|l\rangle e^{-E_lT}\langle l|M^\dagger|0\rangle\nonumber\\
&=&\langle M(T+T')\; M^\dagger(0)\rangle\langle l|{\cal O}|l\rangle.
\label{3point}
\end{eqnarray}
Therefore, one can determine the expectation value of ${\cal O}$ in the
quarkonium state by measuring the ratio of a three-point function to
a two-point function. 

In practice, we rewrite the two-point function in (\ref{3point}) 
as follows: 
\begin{equation}
\lim_{T,T'\rightarrow\infty}\langle M(T+T')\;M^\dagger(0)\rangle ={1\over C}
\langle M(T+T')\;M_{\hbox{p}}(T)\rangle\;
\langle M_{\hbox{p}}^\dagger(T)\;M^\dagger(0)\rangle ,
\label{n-two-point}
\end{equation}
where $M_P$ is a point-source interpolating operator, and $C$ is the
coefficient of $e^{-E_lT}$ in $\langle
M{\hbox{p}}(T)\;M_{\hbox{p}}^\dagger(0)\rangle $.  The sources in the
three-point function match those in the two-point functions on the right
side of (\ref{n-two-point}). Hence, we can reduce noise by measuring the
ratio of the three-point function to those two-point functions. The
expectation value $\langle l|{\cal O}|l\rangle $ is then represented by
the diagram in Fig.~1. 

\begin{figure}[htbp]
\centerline{\begin{picture}(200,125)
\put(50,95){\oval(80,40)}
\put(140,95){\oval(100,40)}
\put(0,65){\line(1,0){200}}
\put(47.5,35){\oval(75,40)}
\put(142.5,35){\oval(95,40)}
\put(10,95){\circle*{10}}
\put(190,95){\circle*{10}}
\put(10,35){\circle*{10}}
\put(190,35){\circle*{10}}
\put(90,95){\circle*{5}}
\put(85,35){\circle*{5}}
\put(95,35){\circle*{5}}
\put(45,95){\vector(-1,0){30}}
\put(55,95){\vector(1,0){32.5}}
\put(135,95){\vector(-1,0){42.5}}
\put(145,95){\vector(1,0){40}}
\put(47,92){$T$}
\put(137,92){$T'$}
\put(-30,62){$C\; \times\;$}
\end{picture}}
\fcaption{Diagrammtic representation of $\langle l|{\cal O}|l\rangle $.  
The large disks represent quarkonium sources (and sinks); the small 
disks represent quarkonium point sources and four-fermion operators.}
\end{figure}

\subsection{Computational method}
\noindent

We measured operator expectation values using noisy-point and
noisy-gaussian sources and generating retarded and advanced quark
propagators from each time slice. We chose the Coulomb gauge for the
field configurations. This gauge choice made the implementation of
extended sources simpler, and it allowed us to replace covariant
derivatives with normal derivatives, with errors of relative order
$v^2$. For matrix elements of covariant operators, we checked some of
our results on non-gauge-fixed field configurations. 

In calculating heavy-quark propagators $G({\bf x},t)$ on the
lattice, we used the nonrelativistic formulation of Lepage, L.~Magnea, 
Nakhleh, U.~Magnea, Hornbostel.\cite{lepage}  We chose an evolution 
equation that is valid to the lowest
non-trivial order in $v^2$, that is, a lattice version of the 
inhomogeneous Schr\"odinger equation:
\begin{equation}
G({\bf x},t+1)=(1-H_0/2n)^n U_{{\bf x},t}^{\dagger} (1-H_0/2n)^n G({\bf x},t)
+\delta_{\bf x,0}\delta_{t+1,0},
\label{evolution}
\end{equation}
with the initial condition $G({\bf x},t)=0$ for $t < 0$. Here, $H_0 =
-\nabla^{(2)}/(2 M_0) - h_0$, $\nabla^{(2)}$ is the gauge-covariant
discrete Laplacian, $M_0$ is the bare heavy-quark mass,
$u_0=\langle(1/3){\rm Tr}\,U_{\rm plaquette}\rangle^\frac{1}{4}$ is the
mean-field value of a gauge-field link, and $h_0=3(1-u_0)/M_0$ is the
mean-field energy shift. This choice of $h_0$ is equivalent at leading
order in $v$ to tadpole improvement of the action.\cite{lepage}  We
chose $n=2$.  For our choices of the bare masses $M_0$, this is the
minimum value of $n$ for which the $Q\bar Q$ propagators are free from
lattice-artifact singularities. 

Note that ${\cal F}_1/M_Q^2$ is suppressed by ${\cal O}(v^2)$ relative
to ${\cal G}_1$.  That is, this ratio is of the same order as terms that
we have neglected in the evolution equation. However, in $^3S_1
\rightarrow \hbox{Light Hadrons}$, $^3S_1 \rightarrow \gamma +
\hbox{Light Hadrons}$, and $^3S_1 \rightarrow 3\gamma$, the coefficient
of ${\cal F}_1/M_Q^2$ is approximately $-5$ times that of ${\cal 
G}_1$.\cite{bbl,bbl3}
Thus, we feel that there is some merit in calculating ${\cal
F}_1/M_Q^2$, even in the presence of order $v^2$ errors. 

\subsection{Parameters of the lattice simulation}
\noindent

In our bottomonium and charmonium measurements, we used of 158 quenched
gauge-field configurations on a $16^3 \times 32$ lattice at $6/g^2=5.7$.
In the case of bottomonium, we also used 149 configurations on a $16^3
\times 32$ lattice at $6/g^2=6.0$. We took our values of the bare
heavy-quark masses from the determination by the NRQCD
collaboration:\cite{nrqcd} $M_{b0}=1.5$ at $6/g^2=6.0$, $M_{b0}=2.7$ at
$6/g^2=5.7$, and $M_{c0}=0.69$ at $6/g^2=5.7$. (Note that our definition
of $M_0$ is $u_0$ times that of the NRQCD collaboration.) We also used
$u_0=0.87778701$ at $6/g^2=6.0$ and $u_0=0.8608261760$ at $6/g^2=5.7$. 

In converting from lattice to physical units, we used the lattice
spacings determined by the NRQCD collaboration.  Their values are
$a^{-1}=2.4$~GeV for bottomonium at $6/g^2=6.0$, $a^{-1}=1.37$~GeV for
bottomonium at $6/g^2=5.7$, and $a^{-1}=1.23$~GeV for charmonium at
$6/g^2=5.7$.  At a given coupling, the lattice spacings are different
for charmonium and bottomonium because the quenched approximation leads
to slight inconsistencies when one tries to use the physical spectra to
fix the lattice spacing. 

\subsection{Lattice results}
\noindent

Our measurements revealed that the vacuum-saturation approximation is
even more accurate than one would expect. For bottomonium at
$6/g^2=6.0$, corrections to the vacuum-saturation approximation for
${\cal G}_1$ are of relative size $1.3(1) \times 10^{-3}$. For
charmonium at $6/g^2=5.7$, the corrections are about $1\%$. The
corrections to the vacuum-saturation approximation for ${\cal H}_1$ are
also small. We assumed that the vacuum-saturation approximation is 
accurate for ${\cal F}_{1}$ as well. The numerical results that we
present for ${\cal G}_{1}$, ${\cal F}_{1}$, and ${\cal H}_{1}$ 
are the vacuum-saturation values. 

The results of our lattice measurements of the matrix elements are shown 
in Table~1. 
\begin{table}[htbp]
\tcaption{Lattice values of the NRQCD matrix elements.}
\centerline{\footnotesize\smalllineskip
\begin{tabular}{|l|c|c|c|}
\hline\hline
\hbox{}& charmonium & \multicolumn{2}{c@{\hspace{8mm}}|}{bottomonium} \\
\hline
$6/g^2$                  & 5.7        & 5.7            & 6.0             \\
\hline
${\cal G}_{1L}$                      & 0.1317(2)(12) & 0.9156(9)(65) 
                                                     & 0.1489(5)(12) \\ 
${\cal F}_{1L}(non)/{\cal G}_{1L}$   & 1.2543(7)     & 2.7456(8) 
                                                     & 1.3135(8)     \\
${\cal F}_{1L}(cov)/{\cal G}_{1L}$   & 0.5950(5)     & 2.1547(7)
                                                     & 0.8522(5)     \\
${\cal F}_{1L}(non_2)/{\cal G}_{1L}$ & 0.7534(4)     & 1.2205(2) 
                                                     & 0.7775(5)     \\  
${\cal F}_{1L}(cov_2)/{\cal G}_{1L}$ & 0.5201(3)     & 1.1111(2) 
                                                     & 0.6659(3)     \\
${\cal H}_{1L}$                      & 0.0208(2)(20) &   -----     
                                                     & 0.0145(6)(20) \\
${\cal H}_{8L}/{\cal H}_{1L}$        & 0.034(2)(8)   &   -----
& 0.0152(3)(20) \\
\hline\hline
\end{tabular}
}
\end{table}
(The subscript $L$ denotes lattice regularization.) The first error in
each quantity is statistical.  Where two errors are given, the second
error is an estimate of systematics associated with the parametrization
of the functions used to fit to the propagators and with contamination
from higher states. The arguments of ${\cal F}_{1L}$ indicate different
lattice representations of the operator. The argument $cov$ denotes a
tadpole-improved naive gauge-covariant operator; the argument $non$
denotes the simple, gauge-noncovariant, finite-difference operator in
the Coulomb gauge. The subscript $2$ indicates a difference operator
with spacings of two lattice units. This softer lattice Laplacian was
useful in controlling regulator-artifact power divergences in the
operator matrix element, which we shall discuss later. 

\vspace*{1pt}\textlineskip      
\section{Relation of Lattice Matrix Elements to Continuum Matrix Elements}
\vspace*{-0.5pt}
\noindent

At leading order in $v^2$, the lattice and continuum matrix elements of 
the operators that we measured are related as follows:
\begin{eqnarray}
{\cal G}_{1L} &=& (1+\epsilon) {\cal G}_1,\nonumber \\
{\cal F}_{1L} &=& (1+\gamma ) {\cal F}_1 + \phi {\cal 
G}_1,\nonumber\\
{\cal H}_{1L} &=& (1+\iota ) {\cal H}_1 + \kappa {\cal H}_8,\nonumber \\
{\cal H}_{8L} &=& (1+\eta) {\cal H}_8 + \zeta {\cal H}_1.
\label{mix}
\end{eqnarray}
(The continuum-regulated matrix elements have no subscript.)

Note that, to leading order in $v^2$, ${\cal G}_{1L}$ has no ${\cal
F}_1$ component.  In fact, if one were to try to compute the addmixture
of ${\cal F}_1$, using the leading order NRQCD action that corresponds
to (\ref{evolution}), then the resulting inconsistencies in the
treatment of $v^2$ corrections would lead to uncanceled IR divergences. 

The coefficients in (\ref{mix}) relate different UV regularizations of
the operators.  Therefore, they are short-distance quantities and,
hence, are perturbatively calculable. The coefficients $\epsilon$,
$\gamma$, $\phi$, $\iota$, $\eta$ and $\zeta$ are of order $\alpha_s$,
while $\kappa$ is of order $\alpha_s^3$. 

We calculated these coefficients through order $\alpha_s$ (one loop) in
tadpole-improved perturbation theory.\cite{lm} First, we obtained analytic
expressions for the integrands.  By carrying out the time components of
the loop integrations analytically, we were able to identify the IR
divergent pieces, which are identical in the lattice and continuum
matrix elements.  After subtracting these divergent pieces, we evaluated
the remaining IR-finite integrals numerically using VEGAS. 

Our numerical results (in lattice units) for the case of
$\overline{MS}$~regularization of the continuum matrix elements are
shown in Table~2. 
\noindent
\begin{table}[htbp]
\tcaption{Coefficients relating lattice and continuum ($\overline{MS}$) 
matrix elements.}
\centerline{\footnotesize\smalllineskip
\begin{tabular}{|l|c|c|c|}
\hline\hline
         & charmonium & \multicolumn{2}{c@{\hspace{1cm}}|}{bottomonium} \\
\hline
$6/g^2$         & 5.7        & 5.7            & 6.0             \\
\hline
$\epsilon$      & -0.7326 $\alpha_s$ & 0.2983   $\alpha_s$ 
                & -0.4877 $\alpha_s$  \\
$\gamma(non)$   & -0.02578 $\alpha_s$ & -1.248   $\alpha_s$ 
                & -0.9117 $\alpha_s$  \\
$\gamma(cov)$   & -2.860   $\alpha_s$ & -2.192   $\alpha_s$ 
                & -2.560  $\alpha_s$  \\
$\gamma(non_2)$ & -0.2774 $\alpha_s$ & -1.096   $\alpha_s$ 
                & -0.9236 $\alpha_s$  \\
$\phi(non)$     & 1.486   $\alpha_s$ & 10.90    $\alpha_s$ 
                & 4.418   $\alpha_s$  \\
$\phi(cov)$     & 0.3928  $\alpha_s$ & 9.808    $\alpha_s$ 
                & 3.325    $\alpha_s$  \\
$\phi(non_2)$   & 1.004  $\alpha_s$ & 6.096   $\alpha_s$ 
                & 2.863  $\alpha_s$  \\
$\iota$         & -0.7603 $\alpha_s$ & -1.852   $\alpha_s$ 
                & -1.191  $\alpha_s$  \\
$\eta$          & 0.09157 $\alpha_s$ & -0.03728 $\alpha_s$ 
                & 0.06096 $\alpha_s$  \\
$\zeta$         & -0.1785 $\alpha_s$ & -0.006011 $\alpha_s$ 
                & -0.01862 $\alpha_s$\\
\hline\hline
\end{tabular}
}
\end{table}
The accuracy of the coefficients of $\alpha_s$ is better than 1\%. The
quantity $\zeta$ depends at one-loop order on the factorization scale,
which we took to be 1.3~GeV for charmonium and 4.3~GeV for bottomonium.
These values are approximately equal to the $\overline {MS}$ c-quark and
b-quark masses, respectively. 

Some of the coefficients of $\alpha_s$ in the expressions for $\phi$
appear to be large.  However, in physical units, ${\cal G}_1$ has
dimensions $({\rm mass})^3$ and ${\cal F}_1$ has dimensions $({\rm
mass})^5$. Hence, we see from (\ref{mix}) that $\phi$ has dimensions of
$({\rm mass})^2$. We can make $\phi$ dimensionless by dividing ${\cal
F}_1$ and $\phi$ by $M_Q^2$.  Similarly, we can make $\kappa$ and
$\zeta$ dimensionless by dividing ${\cal H}_1$ and $\kappa$ by $M_Q^2$
and multiplying $\zeta$ by $M_Q^2$. Using the values, 
$M_b=5.0\hbox{~GeV}$ and $M_c=1.5\hbox{~GeV}$, we find that none of the
dimensionless coefficients of $\alpha_s$ is large. We conclude that the
perturbation series is reasonably behaved in one-loop order. 

In setting the scale for $\alpha_s$, we made use of the method of
Brodsky, Lepage and Mackenzie.\cite{lm,blm}  For coefficients that arise
from a positive integrand, the scale is about $1/a$.  In the case of
integrands without definite sign, large cancellations can make the
normalizing integral anomalously small and spoil the simplest
scale-setting method.  Therefore, we chose the scale to be $1/a$ for all
of the coefficients, taking $\alpha_s=\alpha_V (1/a)=0.3552$ at
$6/g^2=5.7$ and $\alpha_s=\alpha_V (1/a)=0.2467$ at $6/g^2=6.0$. 

\vspace*{1pt}\textlineskip
\section{Continuum Matrix Elements}
\vspace*{-0.5pt}
\noindent

Substituting the lattice matrix elements and the lattice-to-continuum
coefficients into (\ref{mix}), we obtain the results for the
continuum-regulated ($\overline {MS}$) matrix elements shown in the
first two columns of Table~3. 
\begin{table}[htbp]
\tcaption{Continuum-regulated ($\overline {MS}$) matrix elements.}
\centerline{\footnotesize\smalllineskip
\begin{tabular}{|l|c|c|c|}
\hline\hline
             & \multicolumn{2}{c@{\hspace{1.6cm}}|}{lattice} & experiment  \\
\hline
                         & lattice units       & physical units &          \\
\hline
\multicolumn{2}{|l|}%
{charmonium $6/g^2=5.7$}   &                &          \\ 
${\cal G}_1$                  & 0.1780(3)(16)$(^{+366}_{-259})$
                              & 0.3312(6)(30)$(^{+681}_{-483})$~GeV$^3$
                              & 0.36(3)~GeV$^3$                            \\
${\cal F}_1 / {\cal G}_1$       & 0.05 --- 0.54 
                              & 0.07 --- 0.82~GeV$^2$
                              & 0.057~GeV$^2$                              \\
${\cal H}_1$                  & 0.0285(2)(27)$(^{+60}_{-42})$
                              & 0.0802(6)(77)$(^{+167}_{-118})$~GeV$^5$
                              & 0.077(19)(28)~GeV$^5$                       \\
${\cal H}_8 / {\cal H}_1$     & 0.086(1)(6)$(^{+42}_{-32})$
                              & 0.057(1)(4)$(^{+27}_{-21})$~GeV$^{-2}$
                              & 0.095(31)(34)~GeV$^{-2}$                    \\
\hline
\multicolumn{2}{|l|}%
{bottomonium $6/g^2=5.7$}    &                &        \\
${\cal G}_1$                  & 0.8279(8)(59)$(^{+1066}_{-848})$
                              & 2.129(2)(15)$(^{+274}_{-218})$~GeV$^3$
                              & 3.55(8)~GeV$^3$                            \\
${\cal F}_1 / {\cal G}_1$     & -3.7 --- 0.2
                              & -6.9 --- 0.4~GeV$^2$
                              &     -----                                  \\
\hline
\multicolumn{2}{|l|}%
{bottomonium $6/g^2=6.0$}    &                &        \\
${\cal G}_1$                  & 0.1692(6)(14)$(^{+126}_{-110})$
                              & 2.340(8)(19)$(^{+173}_{-151})$~GeV$^3$
                              & 3.55(8)~GeV$^3$                            \\
${\cal F}_1 / {\cal G}_1$     & -0.34 --- 0.28 
                              & -2.0 --- 1.6~GeV$^2$
                              &     -----                                  \\
${\cal H}_1$                  & 0.0205(9)(28)$(^{+23}_{-19})$
                              & 1.63(7)(23)$(^{+19}_{-15})$~GeV$^5$
                              &     -----                                  \\
${\cal H}_8 / {\cal H}_1$     & 0.0151(2)(14)$(^{+33}_{-29})$
                              & 0.00262(3)(24)$(^{+57}_{-51})$GeV$^{-2}$
                              &     -----                    \\
\hline\hline
\end{tabular}
}
\end{table}
The first two errors are the statistical and systematic errors from the
lattice measurements.  The third error is the systematic error from the
neglect of terms of higher order in $\alpha_s$ in the
lattice-to-continuum coefficients. 

Errors from the omission of terms of higher order in $v^2$ in the
evolution equation and in the operator mixing have not been reported in
the first two columns of Table~3.  We expect these errors to be of order
$10\%$ for bottomonium and $30\%$ for charmonium. In the case of ${\cal
G}_{1L}$, the NRQCD collaboration has given results that are accurate to
next-to-leading order in $v^2$. The weighted averages of the singlet-
and triplet-state values are ${\cal G}_{1L}=0.133(4)$ for charmonium at
$6/g^2=5.7$ and ${\cal G}_{1L}=0.144(4)$ for bottomonium at
$6/g^2=6.0$, which are in good agreement with our results.  This
suggests that the approximate effect of the corrections of higher order
in $v^2$ is to split the values of the matrix elements for a multiplet 
of spin states, without changing their spin average.

The second column of Table~3 does not include errors that arise from
uncertainties in the physical values of $a^{-1}$. We estimate, from the
results of the NRQCD collaboration, that these errors are 7\% for ${\cal
G}_1$ in charmonium, 13\% for ${\cal H}_1$ in charmonium, 13\% for
${\cal G}_1$ in bottomonium, and 23\% for ${\cal H}_1$ in bottomonium. 

In addition, there are errors associated with the quenched approximation, 
for which we have no quantitative estimate.

\subsection{Experimental values of the matrix elements}
\noindent

The third column in Table~3 gives phenomenological results for the
matrix elements. ${\cal G}_1$ was extracted from the measured decay
rates for $J/\psi \rightarrow e^+ e^-$, $\eta_c \rightarrow
\gamma\gamma$ and $\Upsilon \rightarrow e^+ e^-$
(Ref.~{\lowcite\cite{rpp}}), using the expressions in given in
Ref.~{\lowcite\cite{bbl}}. The value for ${\cal F}_1/{\cal G}_1$ for
$J/\psi$ is from the calculation of Ko, Lee and Song.\cite{kls} ${\cal
H}_1$ and ${\cal H}_8/{\cal H}_1$ for $\chi_c$ are from
Ref.~{\lowcite\cite{bbl2}}. There is no published data for $\chi_b$
decays into light hadrons, photons, and/or leptons. To extract the
phenomenological matrix elements for ${\cal G}_1$ and ${\cal H}_1$, we
used the values $M_b(\hbox{pole})=5.0~\hbox{GeV}$
(Ref.~{\lowcite\cite{nrqcd}}), $M_c(\hbox{pole})=1.5~\hbox{GeV}$
(Ref.~{\lowcite\cite{eq}}), $\alpha_s(M_c)=0.243$,
$\alpha_s(M_b)=0.179$, $\alpha(M_c)=1/133.3$, and $\alpha(M_b)=1/132$. 

In the third column of Table~3, the first error is experimental, and the
second, where it is given, is theoretical. Where no theoretical error is
given, it is at least as large as the uncertainty from the neglect of
terms of higher order in $\alpha_s$ in the calculation of the
short-distance coefficients. This uncertainty is of nominal size
$25\%$ for charmonium and $20\%$ for bottomonium. Errors that arises
from uncertainties in the heavy-quark masses have not been reported in the
third column. The NRQCD collaboration quotes an error of 4\% for the
$b$-quark mass, which leads to errors of 8\% in ${\cal G}_1$ and 16\% in
${\cal H}_1$. In the case of the $c$-quark mass there is, as yet, no
reliable determination from a lattice calculation. 

\vspace*{1pt}\textlineskip
\section{Discussion}
\vspace*{-0.5pt}
\noindent

In the case of charmonium, our results for ${\cal G}_1$, ${\cal
H}_1$, and ${\cal H}_8/{\cal H}_1$ are in agreement with experiment,
but the errors are large.  It is interesting to note that this agreement would 
have failed if we hadn't included the lattice-to-continuum corrections 
to the matrix elements. 

The ratio ${\cal F}_1/{\cal G}_1$ is poorly determined, largely because
of uncertainties in the coefficient $\phi$ that gives the mixing of
${\cal F}_1$ into ${\cal G}_1$.  The mixing of ${\cal F}_1$ into ${\cal
G}_1$ is power UV divergent. Since the mixing is UV dominated and begins
at one loop, we expect it to be of order $\alpha_s$. On the other hand,
in the continuum, ${\cal F}_1/(M^2{\cal G}_1)$ is of order $v^2\ll 1$.
Therefore the effect of mixing on the ratio ${\cal F}_1/(M^2{\cal G}_1)$
is of relative order $\alpha_s/v^2$ and is large for both charmonium and
bottomonium. 

Nevertheless, we can conclude that ${\cal F}_1/(M^2{\cal G}_1)$ is no
larger than ${\cal O}(v^2)$, in agreement with the NRQCD scaling
rules.\cite{bbl,lepage} ${\cal F}_1/{\cal G}_1$ is probably positive for
charmonium and negative for bottomonium. Note that, because the
continuum matrix element ${\cal F}$ is gotten by subtracting UV
divergences, it need not be positive. 

For bottomonium, the lattice result for ${\cal G}_1$ is 35 --- 40\%
below the experimental value.  We know from the results of the NRQCD 
collaboration that at least part of this discrepancy is due to the quenched 
approximation.\cite{nrqcd,gpl}  There is good agreement between our 
results at $6/g^2=5.7$ and $6/g^2=6.0$, which confirms the expected 
renormalization-group scaling behavior.

Our results for the P-wave matrix elements for bottomonium can be
translated immediately into predictions for bottomonium decay
rates.\cite{bbl2} The values for individual matrix are probably subject
to large corrections from the quenched approximation, but the ratio of
octet to singlet matrix elements may be less susceptible to this source
of error. 

Aside from quenching, the largest uncertainties in the matrix elements
come from neglect higher-order (in $\alpha_s$) corrections to the
lattice-to-continuum coefficients. One might remedy this situation by
using lattice methods to compute the relations between the lattice matrix
elements and the momentum-subtracted continuum matrix elements
nonperturbatively, as suggested by Martinelli and
Sachrajda.\cite{sachrajda}  The momentum-subtracted matrix elements
could then be converted to $\overline {MS}$ matrix elements 
in continuum perturbation theory.

In the continuum, $\overline {MS}$ regularization of the operator matrix
elements leads to renormalon ambiguities.\cite{sachrajda} These
ambiguities are of the same order in $v^2$ as the matrix elements of
operators of higher dimension. In the case of $H_8$, we expect such
ambiguities to be small, since $H_1$ first mixes with $H_8$ in order
$\alpha_s^3$. Renormalon ambiguities are absent in the case of
hard-cutoff regulators, such as the lattice.  That is, they are an
artifact of the regulator (factorization) scheme that one chooses to
define NRQCD. The consistency of NRQCD as an effective theory guarantees
that regulator-scheme dependence is absent in physical quantities. 
Hence, renormalon ambiguities cancel in decay rates if one computes the
NRQCD short-distance coefficients and the lattice-to-continuum
coefficients to the same order in $\alpha_s$. 

It is interesting that, for both charmonium and bottomonium, the values
of ${\cal H}_8 /{\cal H}_1$ that we obtain are in agreement with a crude
phenomenology.\cite{bbl}  In this phenomenology, one obtains ${\cal
H}_8$ by solving the one-loop evolution equation for ${\cal H}_8$, under
the assumption that ${\cal H}_8$ vanishes below a scale $M_Q v$. The
one-loop evolution of the decay matrix element ${\cal H}_8$ is the same
as for the corresponding production matrix element ${\cal H}'_8$.  This
suggests that ${\cal H}'_8\approx {\cal H}_8$. The production matrix
element ${\cal H}'_8$ can be extracted from CDF data for charmonium
production\cite{cl} and from recent CLEO data.\cite{cleo} Using our
value for ${\cal H}_1$, we obtain ${\cal H}'_8/{\cal
H}_1=0.042(19)\hbox{ GeV}^{-2}$ and ${\cal H}'_8/{\cal
H}_1=0.046(28)\hbox{ GeV}^{-2}$, respectively,  both of which are in
good agreement with our result for ${\cal H}_8/{\cal H}_1$. 

\nonumsection{Acknowledgements}
\noindent

We wish to thank G.~Peter~Lepage, John Sloan and Christine Davies for
informative discussions and for access to some of their unpublished
results. We also thank G.~Peter~Lepage for carrying out simulations to
check our S-wave results. Our calculations were performed on the CRAY
C-90 at NERSC, whose resources were made available to us through the
Energy Research Division of the U.~S. Department of Energy. This work
was supported by the U.S. Department of Energy, Division of High Energy
Physics, Contract W-31-109-ENG-38. S.~K. is supported by KOSEF through
CTP. 

\nonumsection{References}


\noindent
\begin{thebibliography}{000}

\bibitem{bks} G.T.~Bodwin, S.~Kim and D.K.~Sinclair, in {\it 
Lattice~'94}, Proceedings of the International Symposium, Bielefeld, 
Germany, edited by  F.~Karsch {\it et al.} [Nucl.\ Phys.\ 
B (Proc.\ Suppl.) {\bf 42} (1995) 306], Report No.\ hep-lat/9412011; 
Phys.\ Rev.\ Lett.\ {\bf 77} (1996) 2376.

\bibitem{caswell-lepage} W.E.~Caswell and G.P.~Lepage, Phys.\ Lett.\
{\bf 167B} (1986) 437. 

\bibitem{bbl} G.T.~Bodwin, E.~Braaten and G.P.~Lepage, Phys.\ Rev.\ D {\bf 51}
(1995) 1125.

\bibitem{color-singlet-model} For a review of the color-singlet model see 
G.A.~Schuler, CERN Report No.\ CERN-TH.7170/94 (hep-ph/9403387).

\bibitem{lepage} G.P.~Lepage, {\it et al.,} 
Phys.\ Rev.\ D {\bf 46} (1992) 4502.

\bibitem{bbl3} G.T.~Bodwin, E.~Braaten and G.P.~Lepage (unpublished).

\bibitem{nrqcd} C.T.H.~Davies {\it et al.,} Phys.\ Rev.\ D {\bf 50}
(1994) 6963;
Phys.\ Rev.\ Lett.\ {\bf 73} (1994) 2654;
Phys.\ Rev.\ D {\bf 52} (1995) 6519; in {\it Lattice~'95}, Proceedings 
of the International Symposium, Melbourne, Australia, edited by 
T.D.~Kieu {\it et al.} [Nucl.\ Phys.\ B (Proc.\ Suppl.) {\bf 47} 
(1996) 409], Report No.\ hep-lat/9510006; 
{\it ibid.} {\bf 47} (1996) 421, Report No.\ hep-lat/9510052.

\bibitem{lm} G.P.~Lepage and P.B.~Mackenzie, Phys.\ Rev.\ D {\bf 48} (1993)
2250. 

\bibitem{blm} S.J.~Brodksy, G.P.~Lepage, and P.B.~Mackenzie, Phys.\
Rev.\ D {\bf 28}, (1983) 228. 

\bibitem{rpp} Review of Particle Properties, Phys.\ Rev.\ D {\bf 50}
(1994) 1173.

\bibitem{kls} P.~Ko, J.~Lee, and H.S.~Song, Phys.\ Rev.\ D {\bf 53} 
(1996) 1409.

\bibitem{bbl2} G.T.~Bodwin, E.~Braaten and G.P.~Lepage, Phys.\ Rev.\ D 
{\bf 46} (1992) R1914.

\bibitem{eq} E.~J.~Eichten and C.~Quigg, Phys.\ Rev.\ D {\bf 52} (1995) 
1726.

\bibitem{gpl} G.P.~Lepage, private communication.

\bibitem{sachrajda} C.T.~Sachrajda, in {\it Lattice~'95}
(Ref.~{\lowcite\cite{nrqcd}}), p.~100, Report No.\ hep-lat/9509085, and
references therein. 

\bibitem{cl} P.~Cho and A.K.~Leibovich, Phys.\ Rev.\ D {\bf 53} (1996)
6203. 

\bibitem{cleo} R.~Balest {\it et al.,} Phys.\ Rev.\ D {\bf 52} (1995) 2661.

\end{thebibliography}
\end{document}